\documentclass[11pt]{article}
\usepackage{amssymb}

\newtheorem{theorem}{Theorem}[section]

\newtheorem{definition}[theorem]{Definition}

\newtheorem{lemma}[theorem]{Lemma}

\newcommand{\qedsymb}{\hfill{\rule{2mm}{2mm}}}
\newenvironment{prof}[1][]{\begin{trivlist}
\item[\hspace{\labelsep}{\bf\noindent Proof#1:\/}] }{\qedsymb\end{trivlist}}
\newenvironment{profsketch}[1][]{\begin{trivlist}
\item[\hspace{\labelsep}{\bf\noindent Proof sketch#1:\/}] }{\qedsymb\end{trivlist}}

\def\calH{{\cal H}}

\def\ra{\rangle}

\newcommand{\be}{\begin{eqnarray}}
\newcommand{\ee}{\end{eqnarray}}

\newcommand\defeq{\stackrel{def}{=}}

\newcommand\ket[1]{{ |{#1} \rangle }}
\newcommand\bra[1]{{ \langle {#1} | }}

\newcommand{\onote}[1]{}
\newcommand{\jnote}[1]{}

\newcommand{\eps}{\varepsilon}
\renewcommand{\epsilon}{\varepsilon}

\begin{document}

\title{\bf 3-Local Hamiltonian is QMA-complete}

\author{
 Julia Kempe \footnote{CNRS-LRI UMR 8623,
 Universit\'e de Paris-Sud, 91405 Orsay, France and Computer Science Division and Department of Chemistry, UC Berkeley. E-Mail: kempe@lri.fr}
 \and
Oded Regev \footnote{Institute for Advanced Study, Princeton, NJ. E-Mail: odedr@ias.edu.} }

\date{\today}

\maketitle

\begin{abstract}
It has been shown by Kitaev that the {\sc 5-local Hamiltonian} problem is QMA-complete. Here we reduce the locality of
the problem by showing that {\sc 3-local Hamiltonian} is already QMA-complete.
\end{abstract}

\section{Introduction}
Complexity theory is one of the cornerstones of theoretical computer science, formalizing the notion of an {\em
efficient} algorithm (see, e.g., \cite{Papadimitriou:book}). With the advent of quantum computing a plethora of new
complexity classes have entered the field. One of the major challenges for theoretical computer science is to
understand their structure and the interrelation between classical and quantum classes.

A seminal result in classical complexity theory is the celebrated Cook-Levin theorem which states that SAT is
NP-complete. Namely, we are given a set of clauses (disjunctions) over a set of $n$ variables and asked whether there
exists an assignment to the variables that satisfies all clauses. Moreover, the 3SAT problem in which each clause
contains at most three literals is also NP-complete. It turns out that the 2SAT problem (where each clause contains at
most two literals) can be solved in polynomial time (actually, there is even a linear time algorithm). However, the
MAX2SAT problem in which we are given an extra number $d$ and asked whether there exists an assignment that satisfies
at least $d$ clauses is still NP-complete.

In this paper we will be interested in the quantum analogues of the above results. For a good introduction the reader
is referred to a recent survey by Aharonov and Naveh \cite{DoritQMASurvey} and to a book by Kitaev, Shen and Vyalyi
\cite{KitaevBook}. Kitaev defined a quantum analogue of the classical class NP and named it BQNP. Strictly speaking,
this class is the quantum analogue of MA, the probabilistic version of NP, and hence we will call it QMA (as was done
in \cite{DoritQMASurvey}).

QMA is naturally defined as a class of promise problems: A promise problem $L$ is a pair ($L_{yes},L_{no}$) of disjoint
sets of strings corresponding to ``Yes'' and ``No'' instances of the problem. The problem is to determine,
given a string $x \in L_{yes} \cup L_{no}$, whether $x\in L_{yes}$ or $x \in L_{no}$. Let ${\cal B}$ be the Hilbert space of a qubit.
\begin{definition}[QMA] \label{Def:QMA}
Fix $\eps=\eps(|x|)$ such that $2^{-\Omega(|x|)} \le \eps \le \frac{1}{3}$. Then, a promise problem $L \in QMA$ if there
exists a  quantum polynomial time verifier $V$ and a polynomial $p$  such that:
\begin{itemize}
 \item[-] $\forall x \in L_{yes} \quad \exists \ket{\xi} \in {\cal B}^{\otimes p(|x|)} \quad Pr \left( V(\ket{x},\ket{\xi} )=1\right)\geq 1-\eps$
 \item[-] $\forall x \in L_{no} \quad \forall \ket{\xi} \in {\cal B}^{\otimes p(|x|)} \quad Pr \left( V(\ket{x},\ket{\xi})=1\right)\leq \eps$
\end{itemize}
where $Pr\left( V(\ket{x},\ket{\xi})=1\right)$ denotes the probability that $V$ outputs $1$ given $\ket{x}$ and $\ket{\xi}$.
\end{definition}
By using amplification methods, it was shown in \cite{KitaevBook} that for any choice of $\eps$ in the above range the
resulting classes are equivalent. In this paper we will assume that $\eps$ is $2^{-\Omega(|x|)}$.

We would also like to find an analogue of the SAT problem. One natural choice is the {\sc Local Hamiltonian} problem.
As we will see later, this problem is indeed a complete problem for QMA:
\begin{definition}
We say that an operator $H:{\cal B}^{\otimes n} \longrightarrow {\cal B}^{\otimes n}$ on $n$ qubits is a $k$-local
Hamiltonian if $H$ is expressible as $H=\sum_{j=1}^r H_j$ where each term is a Hermitian operator acting on at most $k$
qubits.
\end{definition}
\begin{definition} \label{Def:localHam}
The (promise) problem {\sc $k$-local Hamiltonian} is defined as follows:
\begin{itemize}
\item[-] {\bf Input:} A $k$-local Hamiltonian on $n$-qubits $H=\sum_{j=1}^r H_j$ with $r=poly(n)$. Each $H_j$ has a
bounded operator norm $\| H_j \| \leq poly(n)$ and its entries are specified by $poly(n)$ bits. In addition, we are
given two numbers $a$ and $b$ (with $poly(n)$ precision) such that $b-a > 1/poly(n)$. We are promised that the smallest
eigenvalue of $H$ is either at most $a$ or larger than $b$.
 \item[-] {\bf Output:}

$1$ if $H$ has an eigenvalue not exceeding $a$,

$0$ if all eigenvalues of $H$ are larger than $b$.
\end{itemize}
\end{definition}
We note that the original definition required that $0\le H_j \leq 1$ (i.e., that both $H_j$ and $I-H_j$ are
nonnegative, meaning that they only have nonnegative eigenvalues). However, it is easy to see that the two definitions are equivalent: given $H_j$'s such that $\|H_j\|\le
poly(n)$ for each $j$, normalize $a$, $b$ and all the $H_j$'s by a factor of $1/poly(n)$ such that $\|H_j\| \le
\frac{1}{2}$. Then, add half the identity to each $H_j$ (such that $0\le H_j \leq 1$) and $\frac{r}{2}$ to $a$ and $b$
where $r$ is the number of terms in $H$.

It can be seen that the {\sc $k$-local Hamiltonian} problem is NP-hard for all $k\ge 2$. This was recently shown by
Wocjan and Beth \cite{2LocalNP} (see also \cite{DoritQMASurvey}). One possible proof is to show that for any $k\ge 2$
the problem is at least as hard as MAX-$k$-SAT. The idea is to represent the $n$ variables by $n$ qubits and represent
each clause by a Hamiltonian. Each Hamiltonian acts on the $k$ variables that appear in its clause. It `penalizes' the
assignment which violates the clause by increasing its eigenvalue. Therefore, the lowest eigenvalue of the sum of the Hamiltonians corresponds to
the maximum number of clauses that can be satisfied simultaneously.

However, the only known QMA-completeness result was due to Kitaev which showed that the {\sc 5-local Hamiltonian}
problem is QMA-complete \cite{KitaevBook}. An interesting open question which was already mentioned in
\cite{DoritQMASurvey} is whether the locality $5$ is optimal. Given that classically MAX$2$SAT is NP-complete we might
hope to reduce the locality of the Hamiltonians. Our main theorem is the following:
\begin{theorem}
The problem {\sc 3-local Hamiltonian} is QMA-complete.
\end{theorem}

We note that the {\sc 1-local Hamiltonian} problem can be solved in polynomial time by a classical algorithm and is
therefore unlikely to be QMA-complete. We leave the case of {\sc 2-local Hamiltonian} as an open problem. Finally, we
mention that using the methods of~\cite{DoritAdiabatic} one can show that the {\sc 2-local Hamiltonian} problem is
QMA-complete if we allow higher dimensional systems instead of qubits.

\section{Kitaev's Construction}\label{section_kitaev}

In this section we will recall Kitaev's proof that {\sc $O(\log n)$-local Hamiltonian} is QMA-complete (his proof that
{\sc $5$-local Hamiltonian} is QMA-complete follows by a simple modification and we will mention it later). The proof
begins by showing that {\sc $k$-Local Hamiltonian} is indeed in QMA for any $k=O(\log n)$:

\begin{lemma}[\cite{KitaevBook}]\label{inqma}
The {\sc $k$-local Hamiltonian} problem is in QMA for any $k=O(\log n)$.
\end{lemma}

Then, it is enough to show that any problem $L$ in $QMA$ can be reduced to {\sc $O(\log n)$-local Hamiltonian}. Let
$U_x=V(\ket{x},\cdot )=U_T\cdots U_1$ be a quantum circuit of size $T=poly(|x|)$ operating on $N=poly(|x|)$ qubits.
Notice that the input $x\in L$ is encoded into the circuit. We assume without loss of generality that $T\ge N$ and that
each gate $U_i$ operates on two qubits. Moreover, we assume that initially, the first $m=p(|x|)$ qubits contain the
proof and the remaining ancillary $N-m$ qubits are zero (see Definition \ref{Def:QMA}). Finally, we assume that the
output of the circuit is written in the first computation qubit (i.e., it is 1 if the circuit accepts). The Hamiltonian
$H$ that is constructed operates on a space of $n=N+\log (T+1)$ qubits. The first $N$ qubits represent the computation
and the last $\log (T+1)$ qubits represent the possible values $0,\ldots,T$ for the clock. The Hamiltonian is
constructed of three terms,
\begin{equation}
H = H_{in} + H_{out} + H_{prop}.
\end{equation}
The terms are given by
 \begin{eqnarray}
 H_{in} &=& \sum_{i=m+1}^{N} \ket{1}_i\bra{1}_i \otimes \ket{0}\bra{0} \nonumber \\
 H_{out} &=& \ket{0}_1 \bra{0}_1 \otimes \ket{T}\bra{T}  \nonumber\\
 H_{prop} &=& \sum_{t=1}^{T} H_{prop,t}
 \end{eqnarray}
and
\begin{equation} H_{prop,t} = \frac{1}{2} (I \otimes \ket{t}\bra{t}
                + I \otimes \ket{t-1}\bra{t-1}
                - U_t \otimes \ket{t} \bra{t-1}
                - U_t^\dag \otimes \ket{t-1} \bra{t}
                )
\end{equation}
for $1 \leq t \leq T$ where $\ket{\alpha}_i \bra{\alpha}_i$ is the projection on the subspace in which the $i$'th qubit
is $\ket{\alpha}$. It is understood that the first part of each tensor product acts on the space of the $N$ computation qubits and the second part acts on the clock qubits. $U_t$ and $U_t^\dagger$ in $H_{prop,t}$ act on the same computational qubits as $U_t$ does when it is employed in the verifier's circuit $U_x$.
Intuitively, each Hamiltonian `checks' a certain property by increasing the eigenvalue if the property doesn't hold: The Hamiltonian $H_{in}$ checks that the input of the circuit is correct (i.e., none of the last
$N-m$ computation qubits is $1$), $H_{out}$ checks that the output bit indicates acceptance and $H_{prop}$ checks that
the propagation is according to the circuit. Notice that these Hamiltonians are $O(\log n)$-local since there are $\log
(T+1)=O(\log n)$ clock qubits. The proof is completed by the following lemmas and recalling that $\eps$ is chosen to be
$2^{-\Omega(|x|)}$ so that $\frac{c}{T^3}-\frac{\eps}{T+1}>1/poly(n)$:

\begin{lemma}[\cite{KitaevBook}]
Assume that the circuit $U_x$ accepts with probability more than $1-\eps$ on some input $\ket{\xi,0}$. Then the
Hamiltonian $H$ has an eigenvalue smaller than $\frac{\eps}{T+1}$.
\end{lemma}

\begin{lemma}[\cite{KitaevBook}]\label{kitaevsoundness}
Assume that the circuit $U_x$ accepts with probability less than $\eps$ on all inputs $\ket{\xi,0}$. Then all the
eigenvalues of $H$ are larger than $\frac{c}{T^3}$ for some constant $c$.
\end{lemma}
Although the proof of this lemma will not be used in this paper, we sketch it here for completeness:
\begin{profsketch}
We write $H=H'+H_{prop}$ where $H'$ denotes $H_{in}+H_{out}$. We start by noticing that both $H'$ and $H_{prop}$ are
non-negative Hamiltonians. We can lower bound the smallest non-zero eigenvalue of $H'$ by $1$ since it is the sum of
commuting projections. It can also be shown that the smallest non-zero eigenvalue of $H_{prop}$ is at least
$\Omega(1/T^2)$. This, however, is not enough to prove the lemma since, for example, the null-spaces of $H'$ and
$H_{prop}$ might have a non-trivial intersection (i.e., there exists a non-zero vector in their intersection).

The next step is to show that since the circuit $U_x$ accepts with small probability, the angle between the null-spaces
of $H'$ and $H_{prop}$ is not too small (in particular, this implies that the intersection of the two null-spaces is
trivial). More specifically, we define the angle $\theta$ between the null-spaces of $H'$ and $H_{prop}$ by
$$ \cos \theta = \max | \langle \eta_1 | \eta_2 \rangle| $$
where the maximum is taken over all $\eta_1$ in the null-space of $H'$ and $\eta_2$ in the null-space of $H_{prop}$.
Then, one can prove that $\sin^2 \theta \ge \Omega(1/T)$. Finally, it can be shown that the smallest eigenvalue of
$H=H'+H_{prop}$ can be lower bounded by the smallest eigenvalue among the non-zero eigenvalues of $H'$ and $H_{prop}$
times $2\sin^2 \frac{\theta}{2}$. Hence, we get the lower bound
$$ \Omega(1/T^2) \cdot 2\sin^2 \frac{\theta}{2} $$
which is at least $\frac{c}{T^3}$ for some constant $c>0$.
\end{profsketch}

\section{The Construction}

The result of the previous section can be improved to {\sc $5$-local Hamiltonian} by using a unary representation for
the clock and noting that three clock qubits are enough to identify the current time step (and since two computation
qubits are also required, we get 5-local Hamiltonians). In addition, one has to add a Hamiltonian that penalizes clock
qubits which are `illegal', i.e., that do not represent a legal unary encoding. For more detail, see \cite{KitaevBook}.
In this section, we show how to use the result of the previous section to obtain the {\sc $3$-local Hamiltonian}
result. Our construction follows the ideas of Kitaev's 5-local proof. The main difference is that our Hamiltonians use
only one clock qubit instead of three. This requires another modification, namely, the penalty for illegal clock
representations has to be considerably higher.

According to Lemma \ref{inqma}, {\sc $3$-local Hamiltonian} is in QMA. Hence, it is enough to show that any problem
$L$ in QMA can be reduced to the {\sc 3-local Hamiltonian} problem. We are given a circuit $U_x=U_T\cdots U_1$ as in
the previous section. We construct a Hamiltonian $H$ that operates on a space of $N+T$ qubits. The first $N$ qubits
represent the computation and the last $T$ qubits represent the clock. The Hamiltonian is constructed of four terms,
\begin{equation}
H = H_{in} + H_{out} + H_{prop} + H_{clock}.
\end{equation}

The first three terms check that the input of the circuit is correct, that the output bit indicates acceptance and that
the propagation is according to the circuit. As before, tensor products separate the computation qubits from the clock qubits:
 \begin{eqnarray}
 H_{in} &=& \sum_{i=m+1}^{N} {\ket{1}}_i\bra{1}_i \otimes \ket{0}_1\bra{0}_1 \nonumber \\
 H_{out} &=& \ket{0}_1\bra{0}_1 \otimes \ket{1}_T\bra{1}_T \nonumber \\
 H_{prop} &=& \sum_{t=1}^{T} H_{prop,t}  \\
 H_{prop,t} &=& \frac{1}{2} (I \otimes \ket{10}_{t,t+1} \bra{10}_{t,t+1}
                + I \otimes \ket{10}_{t-1,t} \bra{10}_{t-1,t}
                - U_t \otimes \ket{1}_t \bra{0}_t
                - U_t^\dag \otimes \ket{0}_t \bra{1}_t
                ) \nonumber
 \end{eqnarray}
for $2 \leq t \leq T-1$ and
 \begin{eqnarray}
 H_{prop,1} &=& \frac{1}{2} (I \otimes \ket{10}_{1,2} \bra{10}_{1,2}
                + I \otimes \ket{0}_1 \bra{0}_1
                - U_1 \otimes \ket{1}_1 \bra{0}_1
                - U_1^\dag \otimes \ket{0}_1 \bra{1}_1
                ) \nonumber \\
 H_{prop,T} &=& \frac{1}{2} (I \otimes \ket{1}_T \bra{1}_T
                + I \otimes \ket{10}_{T-1,T} \bra{10}_{T-1,T}
                - U_T \otimes \ket{1}_T \bra{0}_T
                - U_T^\dag \otimes \ket{0}_T \bra{1}_T
                ).\nonumber
 \end{eqnarray}

For any $0\le t\le T$, let $\ket{\widehat{t}}$ denote the state
$$\ket{\underbrace{1\ldots1}_t\underbrace{0\ldots0}_{T-t}}.$$ These are the legal unary representations. The last term
is chosen to give a high penalty to states which do not contain a legal unary representation in the clock qubits:
\begin{equation}
H_{clock} = T^{12} \sum_{1\le i < j \le T} \ket{0 1}_{ij} \bra{0 1}_{ij}
\end{equation}
We denote the sum $H_{in} + H_{prop} + H_{out}$ of the computation related Hamiltonians by $H_{comp}$. Note that $H$ is
a sum of 3-local Hamiltonians of bounded norm which can be specified by a polynomial number of bits, as required by
Definition \ref{Def:localHam}. We note that some of the terms in $H_{prop}$ are negative, but this is allowed by
Definition \ref{Def:localHam}.

\begin{lemma}[Completeness]
Assume that the circuit $U_x$ accepts with probability more than $1-\eps$ on some input $\ket{\xi,0}$. Then $H$ has an
eigenvalue smaller than $\frac{\eps}{T+1}$.
\end{lemma}
\begin{prof}
Consider the vector
\begin{equation}
 \ket{\eta} \defeq \frac{1}{\sqrt{T+1}}
   \sum_{t=0}^T U_t\cdots U_1 \ket{\xi,0} \otimes \ket{\widehat{t}}.
\end{equation}
Then,
\begin{eqnarray}
\bra{\eta}H\ket{\eta} = \bra{\eta}H_{in}\ket{\eta}  + \bra{\eta}H_{prop}\ket{\eta}  + \bra{\eta}H_{clock}\ket{\eta}+
\bra{\eta}H_{out}\ket{\eta}
\end{eqnarray}
and it is easy to see that the first three terms are zero. Moreover, since $U_x$ accepts with probability higher than
$1-\eps$,
\begin{equation}
\bra{\eta}H_{out}\ket{\eta} < \frac{\eps}{T+1}.
\end{equation}
\end{prof}

\begin{lemma}[Soundness]
Assume that the circuit $U_x$ accepts with probability less than $\eps$ on all inputs $\ket{\xi,0}$. Then all the
eigenvalues of $H$ are larger than $\frac{c}{T^3}$ for some constant $c$.
\end{lemma}
\begin{prof}
Let $\calH_{legal}$ denote the subspace spanned by states whose clock qubits represent a unary encoding. The orthogonal
space is denoted by $\calH_{illegal}$. We will use a simple upper bound on the operator norm of $H_{comp}$ given by
\begin{equation}
||H_{comp} || \le ||H_{in}||+||H_{out}||+\sum_{t=0}^T||H_{prop,t}|| \le N+1+2T \le 4T.
\end{equation}

\onote{WRONG -Note that $H_{clock}$ can be seen as a projection on $\calH_{illegal}$ multiplied by a large constant.}

We will show that for any unit vector $\ket{\eta}$, $\bra{\eta}H\ket{\eta} \ge \frac{c}{T^3}$. Write
$\ket{\eta}=\alpha_1 \ket{\eta_1} + \alpha_2 \ket{\eta_2}$ with $\ket{\eta_1}\in \calH_{legal}$, $\ket{\eta_2}\in
\calH_{illegal}$, $|| |\eta_1\ra||=|| |\eta_2\ra||=1$ and $\alpha_1,\alpha_2\in [0,1]$ with $\alpha_1^2+\alpha_2^2=1$.
If $\alpha_2 \ge \frac{1}{T^5}$ then
\begin{equation}
\bra{\eta}H\ket{\eta} \ge \bra{\eta}H_{clock}\ket{\eta} - ||H_{comp}|| \ge \alpha_2^2 \cdot T^{12} - 4T > 1.
\end{equation}
It remains to consider the case $\alpha_2 < \frac{1}{T^5}$. Noting that $H_{clock} \geq 0$ we get:
\begin{eqnarray}
 && \bra{\eta}H\ket{\eta} = \bra{\eta}H_{clock}\ket{\eta} + \bra{\eta}H_{comp}\ket{\eta} \ge \bra{\eta}H_{comp}\ket{\eta}= \nonumber \\
 && \alpha_1^2 \bra{\eta_1}H_{comp}\ket{\eta_1} +
    2 \alpha_1 \alpha_2 \mbox{Re}(\bra{\eta_1}H_{comp}\ket{\eta_2}) + \alpha_2^2 \bra{\eta_2}H_{comp}\ket{\eta_2} = \nonumber \\
 &&  \bra{\eta_1}H_{comp}\ket{\eta_1} - \alpha_2^2 \bra{\eta_1}H_{comp}\ket{\eta_1}
 + 2 \alpha_1 \alpha_2 \mbox{Re}(\bra{\eta_1}H_{comp}\ket{\eta_2})  + \alpha_2^2 \bra{\eta_2}H_{comp}\ket{\eta_2}\ge \nonumber \\
 && \bra{\eta_1}H_{comp}\ket{\eta_1} - \frac{1}{T^{10}} || H_{comp} || - \frac{2}{T^5} || H_{comp} || - \frac{1}{T^{10}} || H_{comp} || \ge \nonumber\\
 && \bra{\eta_1}H_{comp}\ket{\eta_1} - \frac{8}{T^9} - \frac{8}{T^4}
     > \bra{\eta_1}H_{comp}\ket{\eta_1} - \frac{9}{T^4},
\end{eqnarray}
where we used the bound on the operator norm $||H_{comp}||$. Therefore, it is enough to show that for any $\eta \in
\calH_{legal}$, $\bra{\eta}H_{comp}\ket{\eta} \ge \frac{c}{T^3}$. We will show that by using Lemma
\ref{kitaevsoundness}:
\begin{eqnarray}
&& \bra{\eta}H_{comp}\ket{\eta} = \bra{\eta}\Pi H_{comp}\Pi \ket{\eta} = \nonumber\\
&& \bra{\eta}\Pi H_{in}\Pi \ket{\eta} + \bra{\eta}\Pi H_{out}\Pi \ket{\eta} + \sum_{t=1}^{T} \bra{\eta}\Pi
H_{prop,t}\Pi \ket{\eta}
\end{eqnarray}
where $\Pi$ is the projection on the subspace $\calH_{legal}$. We compute $\Pi H_{comp} \Pi$:
 \begin{eqnarray}
 \Pi H_{in} \Pi &=& \sum_{i=m+1}^{N} {\ket{1}}_i \bra{1}_i \otimes \ket{\widehat{0}}\bra{\widehat{0}} \nonumber \\
 \Pi H_{out} \Pi &=& {\ket{0}}_1 \bra{0}_1 \otimes \ket{\widehat{T}}\bra{\widehat{T}} \nonumber \\
 \Pi H_{prop,t} \Pi &=& \frac{1}{2} (I \otimes \ket{\widehat{t}}\bra{\widehat{t}}
                + I \otimes \ket{\widehat{t-1}}\bra{\widehat{t-1}}
                - U_t \otimes \ket{\widehat{t}} \bra{\widehat{t-1}}
                - U_t^\dag \otimes \ket{\widehat{t-1}} \bra{\widehat{t}}
                )
 \end{eqnarray}
for $2 \leq t \leq T-1$ and
 \begin{eqnarray}
 \Pi H_{prop,1} \Pi &=& \frac{1}{2} (I \otimes \ket{\widehat{1}} \bra{\widehat{1}}
                + I \otimes \ket{\widehat{0}}\bra{\widehat{0}}
                - U_1 \otimes \ket{\widehat{1}} \bra{\widehat{0}}
                - U_1^\dag \otimes \ket{\widehat{0}} \bra{\widehat{1}}
                ) \\
 \Pi H_{prop,T} \Pi &=& \frac{1}{2} (I \otimes \ket{\widehat{T}} \bra{\widehat{T}}
                + I \otimes \ket{\widehat{T-1}} \bra{\widehat{T-1}}
                - U_T \otimes \ket{\widehat{T}} \bra{\widehat{T-1}}
                - U_T^\dag \otimes \ket{\widehat{T-1}} \bra{\widehat{T}}
                )
 .\nonumber
 \end{eqnarray}

The Hamiltonian $\Pi H_{comp}\Pi$ acts on the Hilbert space $\calH_{legal}$ whose dimension is $2^N\cdot(T+1)$. The
Hamiltonian presented in Section \ref{section_kitaev} acts on a Hilbert space of the same dimension. In fact, notice
that the two Hamiltonians are equivalent up to a renaming of the basis elements. Therefore, Lemma \ref{kitaevsoundness}
implies that for any $\eta \in \calH_{legal}$, $\bra{\eta}H_{comp}\ket{\eta} \ge \frac{c}{T^3}$ which completes the
proof.
\end{prof}

\onote{time -> clock}

\onote{Talk about the difference between 3-local qubits and 2-local 8-bits}

\onote{Should we briefly mention the 4-local qubits ?}


\section*{Acknowledgments}

We wish to thank Dorit Aharonov and Fr\'ederic Magniez for useful discussions. JK's effort is sponsored by the Defense Advanced Research
Projects Agency (DARPA) and Air Force Laboratory, Air Force Materiel Command, USAF, under agreement number
F30602-01-2-0524. OR's research is supported by NSF grant CCR-9987845.

\end{document}